\documentclass[twocolumn, times]{aastex631}

\usepackage{booktabs}
\usepackage{multirow}
\usepackage{amsmath}
\hypersetup{colorlinks=true, citecolor=blue, urlcolor=blue, linkcolor=blue, allcolors = blue}

\begin{document}

\title{Machine Learning vs. Spectral Energy Distribution Fitting: A Comparative Analysis of Accuracy in Stellar Mass Estimation}


\author{Vahid Asadi~\href{https://orcid.org/0009-0005-8897-2385}{\includegraphics[scale=0.04]{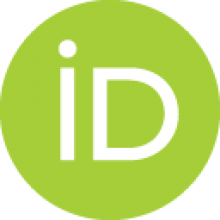}}}
\affiliation{Department of Physics, Institute for Advanced Studies in Basic Sciences (IASBS), PO Box 11365-9161, Zanjan, Iran; \url{vahijd.asadij@gmail.com}}

\author{Akram Hasani Zonoozi~\href{https://orcid.org/0000-0002-0322-9957}{\includegraphics[scale=0.04]{orcid-ID.png}}}
\affiliation{Department of Physics, Institute for Advanced Studies in Basic Sciences (IASBS), PO Box 11365-9161, Zanjan, Iran; \url{vahijd.asadij@gmail.com}}
\affiliation{Helmholtz-Institut f\"ur Strahlen-und Kernphysik (HISKP), Universit\"at Bonn, Nussallee 14-16, D-53115 Bonn, Germany}

\author{Hosein Haghi~\href{https://orcid.org/0000-0002-9058-9677}{\includegraphics[scale=0.04]{orcid-ID.png}}}
\affiliation{Department of Physics, Institute for Advanced Studies in Basic Sciences (IASBS), PO Box 11365-9161, Zanjan, Iran; \url{vahijd.asadij@gmail.com}}
\affiliation{Helmholtz-Institut f\"ur Strahlen-und Kernphysik (HISKP), Universit\"at Bonn, Nussallee 14-16, D-53115 Bonn, Germany}
\affiliation{School of Astronomy, Institute for Research in Fundamental Sciences (IPM), PO Box 19395 - 5531, Tehran, Iran}


\begin{abstract}
Traditional spectral energy distribution (SED)-fitting methods for stellar mass estimation face persistent challenges including systematic biases and computational constraints. We present a controlled comparison of machine learning (ML) and SED-fitting methods, assessing their accuracy, robustness, and computational efficiency. Using a sample of COSMOS-like galaxies from the Horizon-AGN simulation as a benchmark with known true masses, we evaluate the Parametric t-SNE (Pt-SNE) algorithm—trained on noise-injected BC03 models—against the established SED-fitting code LePhare. Our results demonstrate that Pt-SNE achieves superior accuracy, with a root-mean-square error ($\sigma_F$) of 0.169~dex compared to LePhare's 0.306~dex. Crucially, Pt-SNE exhibits significantly lower bias (0.029~dex) compared to LePhare (0.286~dex). Pt-SNE also shows greater robustness across all stellar mass ranges, particularly for low-mass galaxies ($10^9 - 10^{10} M_\odot$), where it reduces errors by 47--53\%. Even when restricted to only six optical bands, Pt-SNE outperforms LePhare using all 26 available photometric bands, underscoring its superior informational efficiency. Computationally, Pt-SNE processes large datasets $\sim 3.2 \times 10^3$ times faster than LePhare. These findings highlight the fundamental advantages of ML methods for stellar mass estimation, demonstrating their potential to deliver more accurate, stable, and scalable measurements for large-scale galaxy surveys.
\end{abstract}

\keywords{Stellar mass estimation --- Machine learning --- SED-fitting --- Pt-SNE --- LePhare}


\section{Introduction}
Stellar mass ($M_{*}$) is a fundamental property of galaxies, governing their evolution and interactions with the cosmic environment \citep[e.g.,][]{bell2001stellar,kauffmann2003dependence,peng2010mass}. Accurate measurements of $M_{*}$ are critical for understanding star formation histories (SFHs), feedback processes, and the assembly of large-scale structures \citep[e.g.,][]{conroy2009propagation,behroozi2013average,somerville2015physical}. While dynamical mass estimates are available for nearby galaxies, their reliance on resolved kinematics makes them challenging for high-redshift objects due to limited angular resolution and faintness. Consequently, most high-redshift studies rely on photometric methods, primarily spectral energy distribution (SED)-fitting \citep[e.g.,][]{boss2000possible,conroy2013modeling}.

Despite their widespread use, traditional SED-fitting tools, such as LePhare \citep{Arnouts1999, Ilbert2006},a widely used and well-documented code, face several well-documented challenges, including inherent uncertainties arising from necessary assumptions about a galaxy's composition \citep[e.g.,][]{Acquaviva2015,Pacifici2015,Iyer2017}, degeneracies between age, dust, and metallicity \citep[e.g.,][]{conroy2009propagation}, sensitivity to assumed SFH parametrizations \citep[e.g.,][]{carnall2018inferring}, and computational inefficiency when applied to large, deep photometric surveys requiring extensive parameter space exploration \citep[e.g.,][]{Hemmati2019,asadi2025leveraging}.

Machine learning (ML) approaches offer promising alternatives by learning non-linear relationships directly from data, helping to break classical degeneracies such as age-dust-metallicity \citep[e.g.,][]{baron2019machine}, achieving orders-of-magnitude speed improvements, and leveraging information from all photometric bands simultaneously \citep[e.g.,][]{Hemmati2019,Davidzon2022,asadi2025leveraging,asadi2025semi,Asadi_2025}. However, a critical consideration for ML models is that if they are trained on simulated data or idealized stellar populations that suffer from the same inherent degeneracies and biases as traditional methods, the ML model will similarly inherit these limitations. 

\cite{asadi2025leveraging} demonstrated that various machine learning algorithms---including Parametric t-SNE \citep[][hereafter Pt-SNE]{Policar2021}, Principal Component Analysis \citep{mackiewicz1993principal}, K-means \citep{macqueen1967some}, Random Forest \citep{Breiman2001}, and Self-Organizing Maps \citep{Kohonen1981}---can achieve stellar mass estimates for COSMOS galaxies \citep{Laigle2016} in the redshift range $0.8 < z < 1.2$ with accuracy comparable to LePhare \citep{Arnouts1999,Ilbert2006} when trained on synthetic simple stellar populations \citep[][hereafter BC03]{Bruzual2003}, while offering significant computational advantages. Their study used LePhare as the benchmark for accuracy comparisons, establishing ML methods as viable alternatives for mass estimation.

Building on this foundation, we present a controlled comparison of ML and SED-fitting methods using a COSMOS-like galaxy catalog derived from the Horizon-AGN hydrodynamical simulation \citep{dubois2014dancing, laigle2019horizon}. The simulated catalog provides true stellar masses as a benchmark, enabling direct validation. We analyze a sample of simulated COSMOS-like galaxies, applying both LePhare (a traditional SED-fitting tool) and Pt-SNE (as a representative ML-based approach), with the latter trained on noise-injected BC03. This approach isolates methodological differences from template systematics while enabling direct validation against ground-truth masses.

Section~\ref{sec:2} describes the data, Sections~\ref{sec:3}--\ref{sec:4} detail the methods, Section~\ref{sec:5} presents the results, and Sections~\ref{sec:6}--\ref{sec:7} discuss the implications and conclude the study.

We adopt a flat $\Lambda$CDM cosmology with parameters $H_{0}=70kms^{-1}Mpc^{-1}$, $\Omega_{m}=0.3$ and $\Omega_{\Lambda}=0.7$. All magnitudes are reported in the AB system \citep{Oke1974}.


\section{Data}\label{sec:2}
\subsection{Mock Galaxies}
This study utilizes a mock galaxy catalog derived from the Horizon-AGN hydrodynamical simulation\footnote{\url{http://www.horizon-simulation.org/}} \citep{dubois2014dancing}. The catalog, presented in \citep{laigle2019horizon}, contains 789,354 galaxies extracted from a $1 \times 1$ deg$^2$ lightcone using the AdaptaHOP halo finder \citep{aubert2004origin} applied to the stellar particle distribution. Each stellar particle (with mass $\sim 2 \times 10^6 M_\odot$) is associated with a synthetic simple stellar population using the BC03 models, assuming a Chabrier initial mass function (IMF) \citep{Chap2003} and interpolating between the metallicity values available in BC03. 

The galaxy sample is selected based on stellar mass ($M_{\rm sim} > 10^9 M_\odot$) and redshift ($0 < z_{\rm sim} < 4$).

The Horizon-AGN virtual observatory reproduces the optical and near-infrared (NIR) photometry of COSMOS2015 galaxies \citep{Laigle2016}, including:
\begin{itemize}
    \item [--] 10 broad bands: u, B, V, r, $i^{+}$, $z^{++}$, $Y$, J, H, $K_s$
    \item [--] 14 medium-band filters (Subaru/SuprimeCam; \citep{taniguchi2007cosmic})
    \item [--] Spitzer/IRAC channels at 3.6 $\mu$m and 4.5 $\mu$m (denoted ch1 and ch2)
\end{itemize}

For each filter, the signal-to-noise ratio (SNR) distribution replicates that of the ultra-deep stripes of COSMOS2015. The reference $3\sigma$ detection limits (measured in 3$''$ apertures) are $K_s < 24.7$ and $i^+ < 26.2$ (complete sensitivity depths for all filters are provided in Table 1 of \citep{Laigle2016}).

Observational uncertainties are incorporated by perturbing the original galaxy fluxes according to their SNR. The model accounts for attenuation by both dust and the inter-galactic medium, though it does not include flux contamination from nebular emission.

\subsection{Sample Selection}
\begin{figure*}
        \centering
        \includegraphics[width=0.9\linewidth]{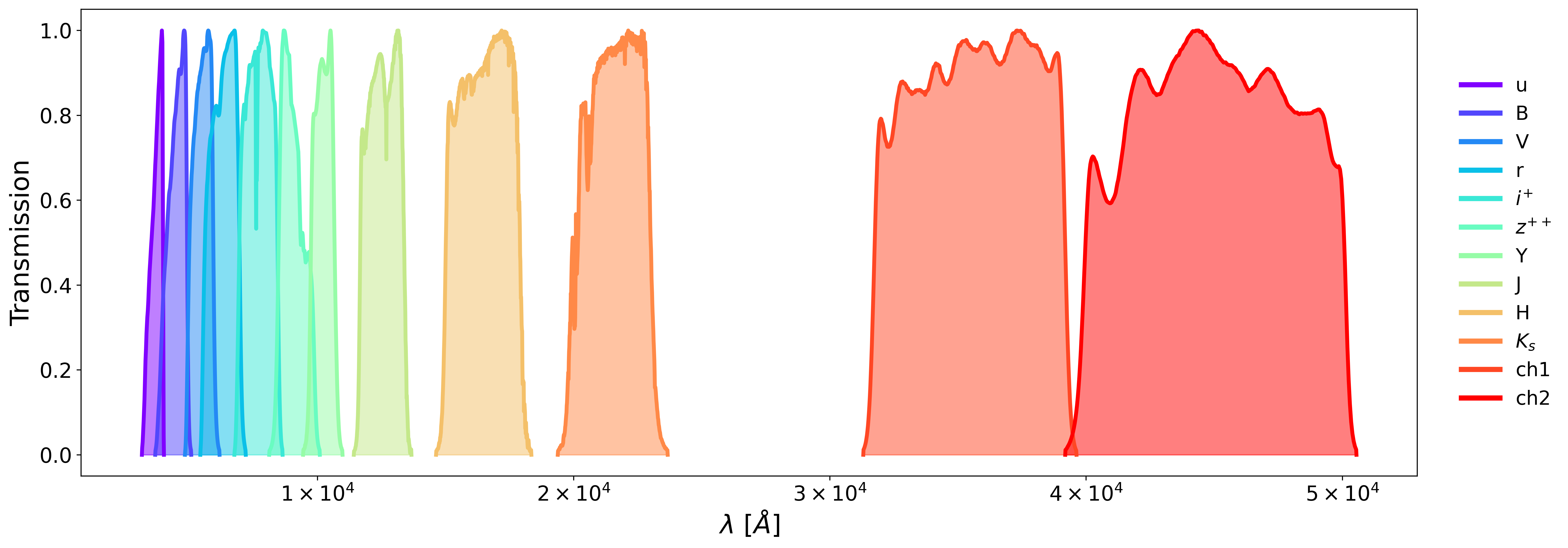}
        \caption{Transmission curves for the photometric bands used for machine learning method.}
        \label{fig:fig_band}
\end{figure*}

The galaxy sample is selected from the Horizon-AGN mock catalog using the following criteria:

\begin{itemize}
    \item [--] Redshift range: $0.8 < z_{\rm sim} < 1.2$
    \item [--] SNR limit: SNR $>$ 1.5 in all photometric bands to exclude non-detections (e.g., when the observed flux is smaller than the flux error),
    \item [--] NIR detection limit: $K_s < 24.7$, matching the depth of the COSMOS2015 ultra-deep stripes.
\end{itemize}
The application of these criteria yields a final sample of 91,261 galaxies. To estimate the stellar masses of this galaxy sample, we use two methods:
\begin{itemize}
    \item [--] Machine Learning Method (Section~\ref{sec:3}) $\rightarrow$ Uses 12 bands: u, B, V, r, $i^{+}$, $z^{++}$, Y, J, H, $K_{s}$, ch1, and ch2 to sample the light from both old and young stars (Figure~\ref{fig:fig_band}),
    \item [--] SED-fitting Method (Section~\ref{sec:4}) $\rightarrow$ Uses all 26 available bands in the mock catalog.
\end{itemize}
This deliberate choice of a reduced band set for the ML method (compared to the 26 bands used for the SED-fitting method) will allow us to investigate ML method's efficiency in extracting information, even from more limited photometric data.


\section{Machine Learning Method}\label{sec:3}
\subsection{Algorithm}
This study employs Pt-SNE\footnote{\url{https://opentsne.readthedocs.io/en/latest/parameters.html}} \citep{Policar2021}, an advanced extension of the t-SNE algorithm \citep{van2008visualizing}, to analyze high-dimensional galaxy data. This method learns a neural network that maps high-dimensional inputs to a low-dimensional embedding space, which then serves as a reduced feature space for predicting physical properties like stellar mass. The algorithm preserves the local neighborhood structure of the data, which is crucial for subsequent property inference. 

Perplexity is a key hyperparameter in this implementation, which controls the effective number of neighbors considered when modeling pairwise similarities and fundamentally influences the resulting embedding structure.

The parametric formulation provides the unique capability to project new galaxy observations into an existing embedding without recomputing the entire visualization. This feature proves particularly valuable for astronomical studies where datasets frequently expand with new observations.

\subsection{Training Data}
To train the Pt-SNE, we utilized BC03 stellar population synthesis code to generate galaxy spectra with different physical properties. The SED of modeled galaxies is constructed by combining simple stellar populations with a range of stellar ages ($7.7<\log_{10}(\rm age/yr)<10.0$), according to an exponentially declining SFH. The e-folding timescale, $\tau$, is adopted to vary in the range of 0.1-10 Gyr. For each simple stellar population, a sub-solar metallicity ($0.4Z_{\odot}$) is adopted and stars are assumed to form following the Chabrier IMF \citep{Chap2003}. Furthermore, we added Calzetti extinction \citep{Calzetti2000} with varying degrees of reddening ($0<E(B-V)<1$) to the synthetic SEDs.

\begin{figure*}
        \centering
        \includegraphics[width=0.85\linewidth]{fig2.png}
        \caption{Manifold of Pt-SNE (black points), with the mapped COSMOS-like mock galaxies (colored points). Galaxy colors correspond to the estimated stellar masses obtained with Pt-SNE. The dimensions of the 2D plot (D1 and D2) are arbitrary labels and do not carry any physical significance.}
        \label{fig:fig_Pt-SNE_mani}
\end{figure*}

To generate synthetic photometry that can be compared to our galaxy mock sample, we adjusted the models to account for a redshift of approximately $z \sim 1$ and integrated them using the same filter transmissions (u,B,V,r,$i^{+}$,$z^{++}$,Y,J,H,$K_{s}$,ch1,ch2) as in the COSMOS-like galaxy sample. As a result, we produced around 14,000 models.

\subsection{Pre-processing}
For training, we use 66 unique colors constructed from the selected bands (Figure~\ref{fig:fig_band}), with pairings designed to capture a broad range of spectral features. These include both adjacent bands (e.g., $u - B$, $B - V$, $V - r$) and widely separated bands (e.g., $u - K_{s}$, $B - ch1$, $i^{+} - J$).

We opted to use color indices rather than apparent magnitudes because they provide a more direct and robust probe of the physical properties of galaxies. Unlike apparent magnitudes, which are sensitive to distance-dependent effects and flux calibration uncertainties, color indices are distance-independent and more resilient to systematic biases. Additionally, because colors measure the relative flux across different wavelengths, they better constrain the shape of the SED, which is crucial for distinguishing between different stellar populations and dust attenuation effects.

To make the idealized BC03 templates more realistically compatible with the COSMOS-like galaxy sample, we mimic the corresponding color noise properties of the COSMOS-like galaxies through a three-step process:

\begin{itemize}
\item [--] First, for each color in the COSMOS-like sample, we trained a \texttt{RandomForestRegressor} model\footnote{\url{https://scikit-learn.org}} \citep[e.g., see][]{macqueen1967some,biau2016random} using the color values as features and their corresponding color errors as targets. 

\item [--] Second, the trained model was applied to predict color errors for the corresponding colors in the BC03 training sample.

\item [--] Third, to simulate COSMOS-like galaxy color errors, we drew random values $\delta F$ from a zero-centered Gaussian distribution scaled by the predicted color errors, and added these perturbations to the training data colors.

\end{itemize}

Before training, we scaled the perturbed training colors using the \texttt{StandardScaler} function\footnote{\url{https://scikit-learn.org}} that performs standardization by transforming each color feature to have zero mean and unit variance:

\begin{equation}
X_{\text{scaled}} = \frac{X - \mu}{\sigma},
\end{equation}

where $\mu$ is the mean and $\sigma$ is the standard deviation of each color feature. This preprocessing step ensures that all color dimensions contribute equally during machine learning by removing arbitrary differences in scale between colors.

\subsection{Training and mass estimation}\label{sec:3.4}
The overall stellar mass estimation process using Pt-SNE consisted of two main stages. First, the Pt-SNE algorithm was trained on the perturbed color data of BC03 galaxies to learn a low-dimensional (2D) embedding space that preserves the local neighborhood structure. After this training, mock COSMOS-like galaxies were then mapped onto this learned 2D manifold. Second, for each mock galaxy mapped onto the manifold, we estimated its stellar mass-to-light ratio ($M/K_{s}$) by computing a distance-weighted mean of the $M/K_{s}$ values of the ten BC03 nearest training data points in the 2D embedding space using Euclidean distance. After assigning this $M/K_{s}$ to each mapped mock galaxy, we obtained its stellar mass by multiplying with its corresponding mock $K_{s}$-band luminosity:

\begin{equation}
M_{*} = \left( \frac{M}{L_{K_{s}}} \right)_{Pt-SNE} \times L_{K_{s}, mock}
\end{equation}

Figure~\ref{fig:fig_Pt-SNE_mani} illustrates the resulting manifold with the mapped mock galaxies.

We adopted Pt-SNE with its default settings (perplexity = 30, a key hyperparameter) after empirical validation demonstrated that hyperparameter optimization yielded only marginal improvements. To evaluate the influence of perplexity, we performed systematic tests by randomly splitting the training model galaxies into a training set (70\%) and a test set (30\%). We varied the perplexity from 10 to 100, trained Pt-SNE on the training set, and computed predictions on the test set. The predicted mass-to-light ratios were compared to the true values using the root-mean-square error (RMS). Our analysis revealed that adjusting the perplexity typically resulted in changes to the RMS of less than 0.01 dex, supporting the adequacy of the default parameters. This finding aligns with modern machine learning practices, where library defaults often provide robust performance across diverse datasets.


\section{SED-fitting Method}\label{sec:4}
Stellar mass estimates for the galaxies were extracted from the Horizon-AGN catalog \citep{laigle2019horizon}. These values were derived by the catalog's creators through a SED-fitting analysis using the LePhare code. The fitting methodology followed the same procedure as the COSMOS2015 catalog \citep{Laigle2016}.

The analysis utilized the BC03 template library model, adopting a Chabrier IMF \citep{Chap2003}. The models considered two different stellar metallicities ($Z_\odot$ and $0.4Z_\odot$) and two extinction laws: \cite{arnouts2013encoding} and \cite{Calzetti2000}. For the star formation histories (SFHs), two analytic parameterizations were employed: Exponentially declining: $\mathrm{SFR}(t) \propto e^{-t/\tau}$ and Delayed (for gas infall modeling): $\mathrm{SFR}(t) \propto te^{-t/\tau}/\tau$. where $\tau$ represents the e-folding time. The following time-steps are used: $\tau$ = 0.1, 0.3, 1, 3, 4, and 30 Gyr for the exponentially declining SFHs, and $\tau$ = 1 and 3 Gyr for the delayed SFHs. In the delayed SFH case, $\tau$ also corresponds to the galaxy age at peak SFR. These models initiate star formation at $t=0$ and do not reproduce multiple bursts. Stellar population ages are sampled up to the age of the Universe at each galaxy's redshift, with up to 44 time steps at $z=0$.


\section{Result}\label{sec:5}

\subsection{Metrics}\label{sec:5.1}
We evaluate the stellar mass estimation performance of Pt-SNE and LePhare by comparing their predicted masses to the true simulation values. Four key statistical metrics are used: the root-mean-square ($\sigma_F$); the normalized median absolute deviation ($\sigma_{\text{NMAD}}$); the standard deviation ($\sigma_{\text{STD}}$); and mean offset (Bias). These metrics are defined as follows:

\begin{equation}
\sigma_{F} = \text{rms}(\log(M_{\text{Sim}}) - \log(M_{\text{Method}}))
\label{eq:rms}
\end{equation}

\begin{equation}
\sigma_{\text{NMAD}} = 1.48 \times \text{median}(|\log(M_{\text{Sim}}) - \log(M_{\text{Method}})|)
\label{eq:nmad}
\end{equation}

\begin{equation}
\sigma_{\text{STD}} = \text{std}(\log(M_{\text{Sim}}) - \log(M_{\text{Method}}))
\label{eq:std}
\end{equation}

\begin{equation}
\text{Bias} = \text{mean}(\log(M_{\text{Sim}}) - \log(M_{\text{Method}}))
\label{eq:bias}
\end{equation}

We identified outliers using the Outlier Fraction (OLF), calculated as the percentage of galaxies where the absolute difference between simulated and predicted log-masses exceeded 0.5 dex. This threshold corresponds to approximately a 3$\sigma$ deviation in log-mass estimates, following the convention established by \citet{mobasher2015critical}. After excluding outliers, we recomputed the root-mean-square error ($\sigma_{0}$) for the cleaned sample.

\begin{table}
\centering
\begin{tabular}{c|cccccccc}
\hline \hline 
\toprule
\multicolumn{1}{c|}{Method} & $\sigma_{F}$ & $\sigma_{NMAD}$ & $\sigma_{STD}$ & OLF & $\sigma_{0}$ & Bias \\ 
\midrule
Pt-SNE  & 0.169 & 0.171 & 0.166 & 0.004 & 0.165 & 0.029\\
\midrule
LePhare & 0.306 & 0.415 & 0.110 & 0.031 & 0.291 & 0.286\\

\bottomrule
\end{tabular}

\caption{Mass prediction metrics comparing Pt-SNE and LePhare against simulations: root-mean-square ($\sigma_F$), normalized median absolute deviation ($\sigma_{NMAD}$), standard deviation ($\sigma_{STD}$), outlier fraction (OLF), outlier-corrected RMS ($\sigma_0$) and mean offset (Bias).}
\label{tab:tab_acc}
\end{table}

\subsection{Overall Comparative Analysis}\label{sec:5.2}
As summarized in Table~\ref{tab:tab_acc}, our analysis reveals a significant difference in the performance characteristics of the Pt-SNE and LePhare methods, with the most striking feature being a systematic bias in the LePhare results.

\begin{figure}
        \centering
        \includegraphics[width=\linewidth]{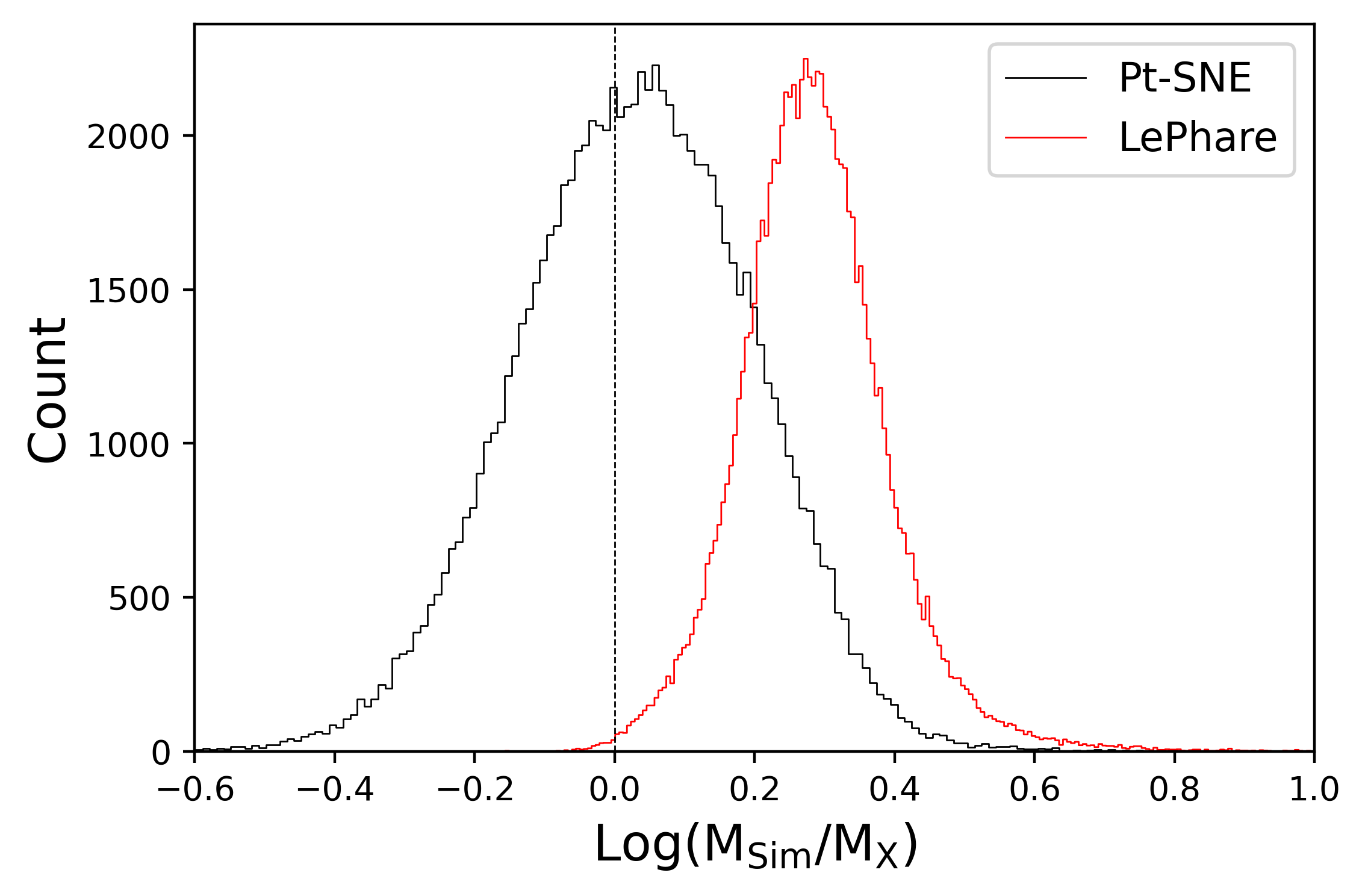}
        \caption{The residual distributions show a significant systematic bias in LePhare (red) versus the unbiased Pt-SNE results (black). LePhare's tighter distribution indicates precision but poor accuracy, while Pt-SNE provides more accurate mass estimates. The dashed line marks perfect agreement.}
        \label{fig:fig_dis}
\end{figure}

The primary advantage of Pt-SNE is its substantially smaller systematic bias (0.029 dex), nearly an order of magnitude lower than LePhare's significant offset (0.286 dex). This indicates that Pt-SNE predictions are well-centered on the true values, while LePhare exhibits a substantial systematic underestimation of stellar masses (see Figure~\ref{fig:fig_dis}).

When considering the random scatter of the residuals, the methods present a trade-off. LePhare achieves a lower standard deviation ($\sigma_{STD} = 0.110$ dex) compared to Pt-SNE (0.166 dex), indicating that its mass estimates are more precise—they are tightly clustered around its own biased mean value. However, this precision is critically undermined by its large systematic error, which severely compromises its overall accuracy.

\begin{figure*}
        \centering
        \includegraphics[width=0.975\linewidth]{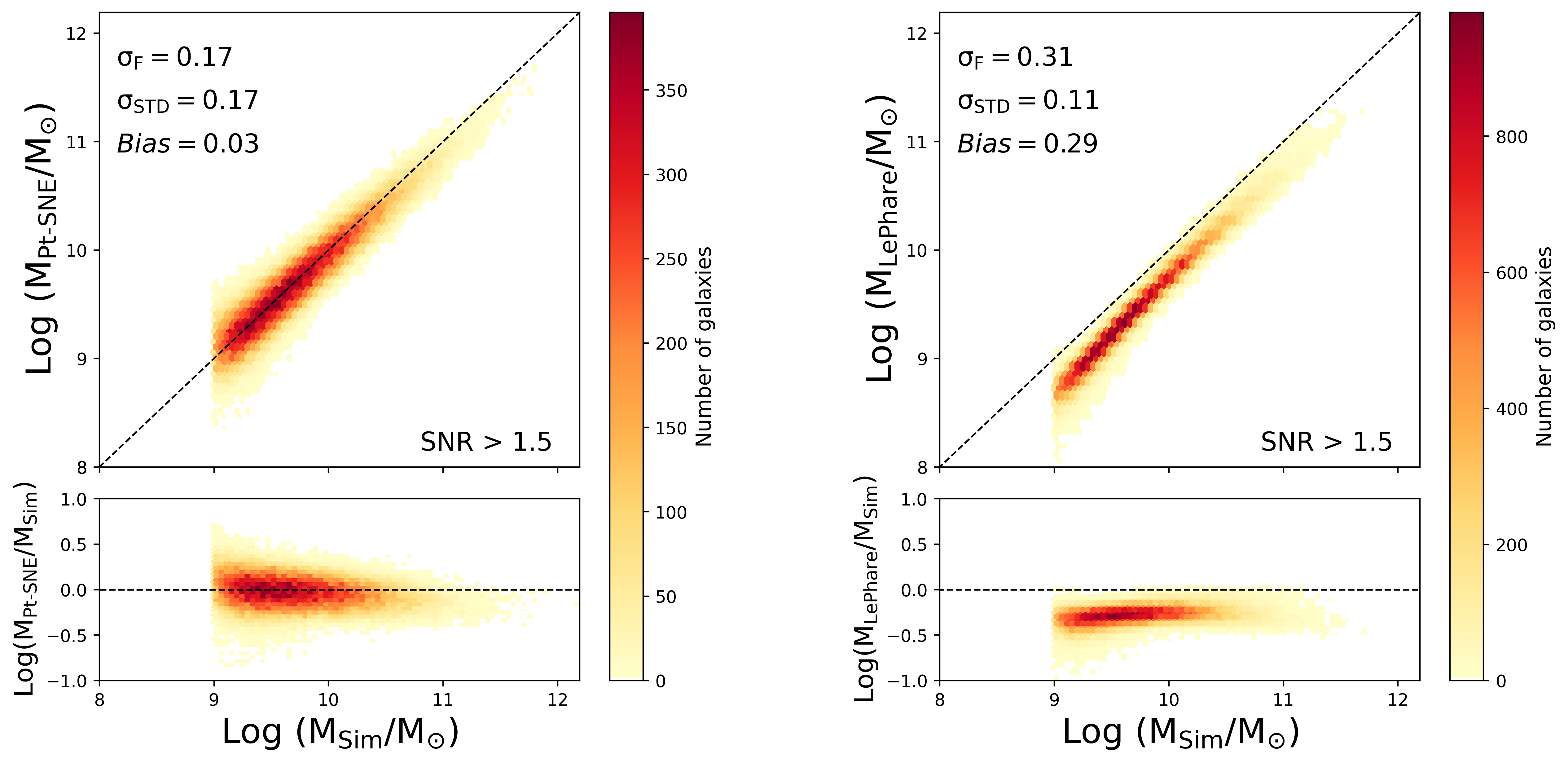}
        \caption{Comparison of the stellar masses of 91,261 COSMOS-like galaxies, derived using Pt-SNE ($M_{Pt-SNE}$) and LePhare ($M_{LePhare}$), with those obtained from simulation ($M_{Sim}$). $\sigma_F$, $\sigma_{STD}$ and Bias are root-mean-square, standard deviation and mean offset respectively.}
        \label{fig:fig_MLvsLePh}
\end{figure*}

In contrast, Pt-SNE demonstrates superior robustness. This is evidenced by its significantly lower normalized median absolute deviation ($\sigma_{NMAD} = 0.171$ dex vs. LePhare's 0.415 dex) and a far lower outlier fraction (OLF = 0.004 vs. 0.031). The minimal difference between Pt-SNE's RMS ($\sigma_F$) and NMAD (0.002 dex) compared to LePhare's larger discrepancy (0.109 dex) further confirms that LePhare produces more catastrophic outliers, while Pt-SNE provides more stable and reliable predictions across the entire sample.

Consequently, Pt-SNE achieves a lower total RMS error ($\sigma_F = 0.169$ dex) compared to LePhare (0.306 dex). After removing outliers, Pt-SNE maintains a lower refined RMS ($\sigma_0 = 0.165$ dex) compared to LePhare ($\sigma_0 = 0.291$ dex). Figure~\ref{fig:fig_MLvsLePh} visually compares the stellar mass estimates from both methods against the simulated values.

The results demonstrate that Pt-SNE provides superior accuracy and robustness compared to LePhare for stellar mass estimation. Importantly, Pt-SNE achieves this enhanced performance using only 12 broad-band filters—compared to the 26 bands used in LePhare’s SED-fitting. This highlights the model’s ability to efficiently extract meaningful information from a reduced photometric dataset, underscoring the potential of ML techniques for large photometric surveys.

\subsection{Stellar mass-Dependent Performance}\label{sec:5.3}
\begin{figure*}
        \centering
        \includegraphics[width=0.91\linewidth]{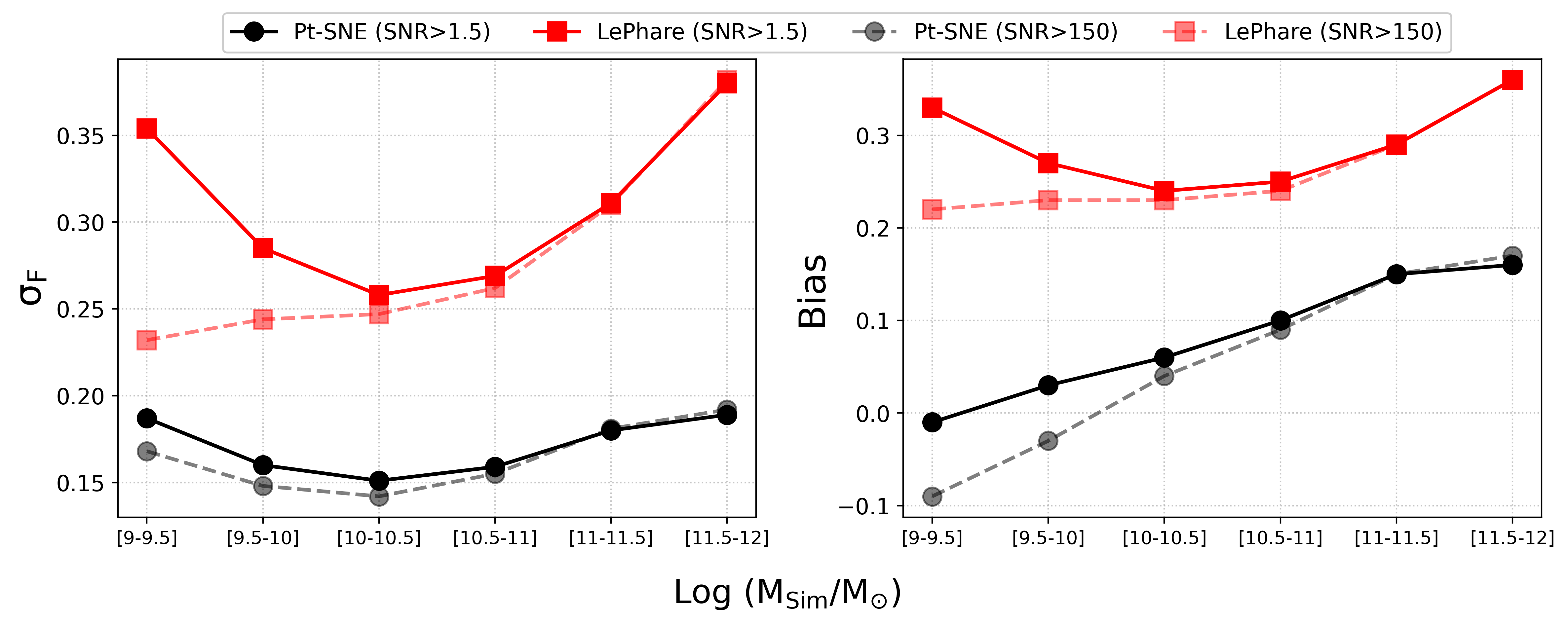}
        \caption{Error trends across simulation-based mass bins ($\log(M/M_{\odot})$) for SNR $> 1.5$ (low-SNR) and SNR $> 150$ (high-SNR). Metrics shown are the root-mean-square error ($\sigma_F$, Equation~\ref{eq:rms}) and bias (Bias, Equation~\ref{eq:bias})}
        \label{fig:fig_mass_bin}
\end{figure*}

We investigated how estimation accuracy varies with stellar mass by computing $\sigma_{F}$ and bias across six logarithmically-spaced mass bins in $\log(M_{Sim}/M_{\odot})$, each with a 0.5 dex width, spanning $\log(M/M_{\odot}) = 9$--$12$. The analysis was performed for both Pt-SNE and LePhare under two different SNR regimes: SNR $> 1.5$ (low signal-to-noise) and SNR $> 150$ (high signal-to-noise). Figure~\ref{fig:fig_mass_bin} visually summarizes these trends.

\begin{table*}
\centering
\begin{tabular}{c|cccccccc}
\hline \hline 
\toprule
\multicolumn{1}{c|}{Method} & Number of Bands & $\sigma_{F}$ & $\sigma_{NMAD}$ & $\sigma_{STD}$ & OLF & $\sigma_{0}$ & Bias \\ 
\midrule
        & 12 (All considered bands) & 0.169 & 0.171 & 0.166 & 0.004 & 0.165 & 0.029\\
        & 10 (Excl. ch1 \& ch2) & 0.196 & 0.192 & 0.186 & 0.015 & 0.184 & 0.061\\
Pt-SNE  & 8 (Excl. Y,J,H,$K_{s}$) & 0.181 & 0.177 & 0.173 & 0.010 & 0.170 & 0.052\\
        & 6 (Optical only) & 0.208 & 0.181 & 0.202 & 0.026 & 0.175 & 0.053\\
        & 6 (Non-optical) & 0.293 & 0.257 & 0.263 & 0.093 & 0.224 & 0.134\\
\midrule
LePhare & 26 (All catalog bands) & 0.306 & 0.415 & 0.110 & 0.031 & 0.291 & 0.286\\

\bottomrule
\end{tabular}

\caption{Comparison of mass prediction metrics between Pt-SNE and LePhare as a function of photometric bands. Root-mean-square ($\sigma_{F}$), normalized median absolute deviation ($\sigma_{NMAD}$), standard deviation ($\sigma_{STD}$), outlier fraction (OLF), outlier-corrected RMS ($\sigma_0$) and Bias.}
\label{tab:tab_acc_band}
\end{table*}

Across the entire mass range, LePhare exhibits significant systematic bias that follows a distinct mass-dependent pattern. The bias is most severe at the mass extremes: at low masses ($\log(M/M_{\odot}) = 9$--$9.5$), LePhare underestimates masses by 0.33 dex at low SNR, while at high masses ($\log(M/M_{\odot}) = 11.5$--$12$) the bias reaches 0.36 dex. The bias minimizes for intermediate-mass galaxies ($\log(M/M_{\odot}) = 10$--$10.5$) at 0.24 dex. This non-uniform, U-shaped bias profile suggests the systematic errors are tied to galaxy physical properties and the limitations of SED-fitting in different regimes, rather than a simple uniform offset that would affect all masses equally.

In contrast, Pt-SNE maintains minimal bias across most of the mass range ($<$0.10 dex for $\log(M/M_{\odot}) < 10.5$), with only a slight positive bias emerging at the highest mass bins. The stability of Pt-SNE's bias profile underscores its robustness across diverse galaxy populations.

The RMS errors ($\sigma_{F}$) reflect this fundamental difference in bias characteristics. Pt-SNE maintains lower $\sigma_{F}$ than LePhare across all mass bins, with the advantage particularly pronounced at lower masses ($\log(M/M_{\odot}) = 9$--$10$), where Pt-SNE shows 47-53\% lower errors at SNR$>$1.5 (0.187-0.160 dex versus LePhare's 0.354-0.285 dex). Both methods exhibit their best performance for intermediate masses ($\log(M/M_{\odot}) = 10$--$10.5$), though Pt-SNE achieves substantially better accuracy ($\sigma_F = 0.151-0.159$ dex at SNR$>$1.5 compared to LePhare's 0.258-0.285 dex).

The SNR analysis reveals important methodological differences in both scatter and bias. While both methods show improved $\sigma_F$ at high SNR for low-mass galaxies, LePhare exhibits much greater sensitivity to noise: its $\sigma_{F}$ decreases by 34\% (0.354 $\rightarrow$ 0.232~dex) with increased SNR, compared to Pt-SNE's 10\% improvement (0.187 $\rightarrow$ 0.168~dex). More notably, LePhare's bias at low masses improves significantly with higher SNR (0.33 $\rightarrow$ 0.22~dex), whereas Pt-SNE's minimal bias remains stable. This pattern suggests that LePhare's performance is more strongly limited by observational noise in the fainter, low-mass regime, whereas Pt-SNE maintains more robust performance even under low-SNR conditions. This SNR sensitivity gradually diminishes with increasing mass, disappearing above $\log(M/M_{\odot}) = 10.5$ for both scatter and bias metrics.

The mass-dependent performance profiles demonstrate that Pt-SNE provides more accurate and stable stellar mass estimates across the entire mass spectrum, while LePhare's significant and variable systematic bias—peaking at both low and high masses—compromises its reliability.

\begin{figure*}
        \centering
        \includegraphics[width=0.8\linewidth]{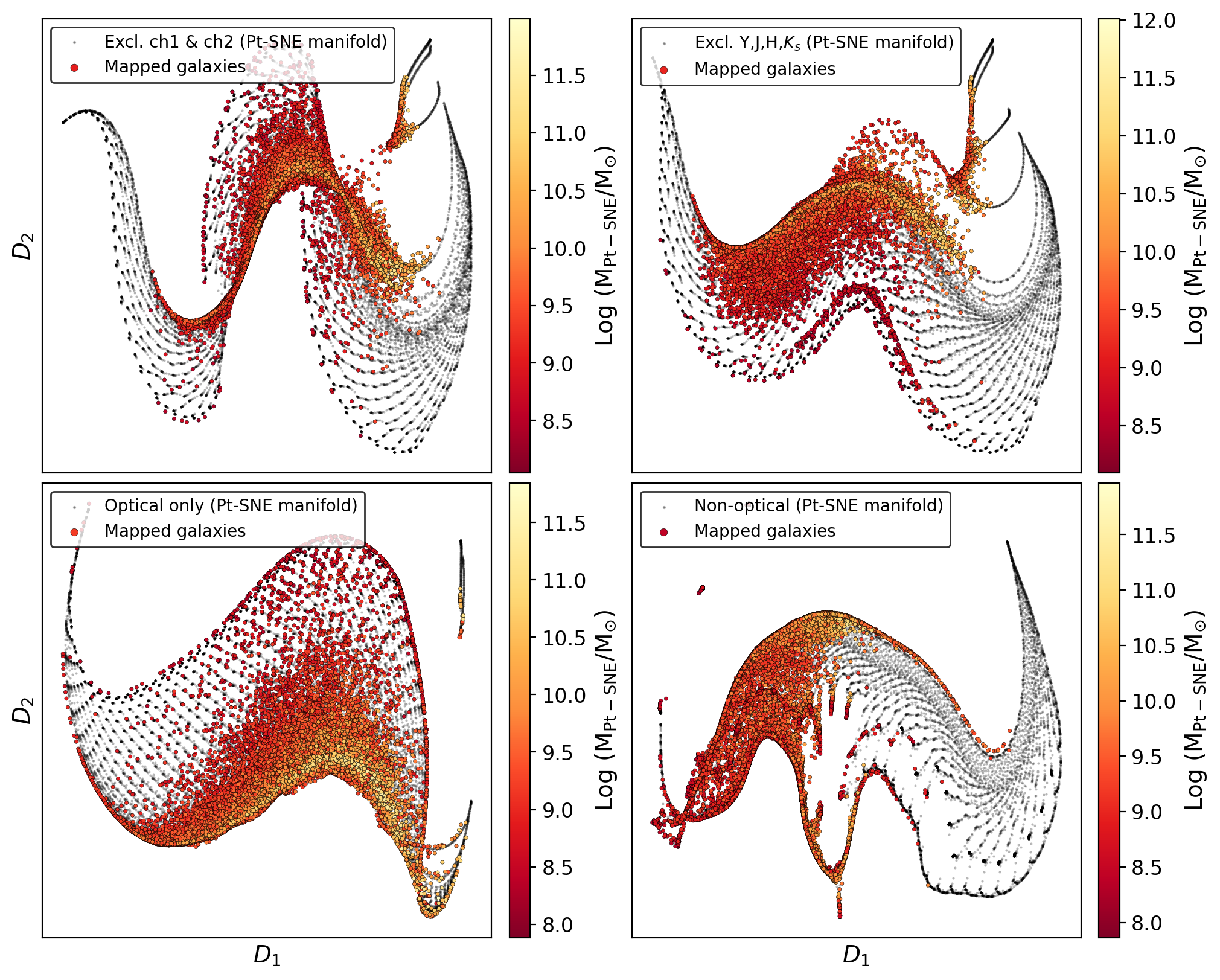}
        \caption{Manifold visualizations of Pt-SNE for four different band configurations (black points show the band-set structure, colored points represent mock COSMOS-like galaxies). Colors correspond to estimated stellar masses, demonstrating how different band combinations affect the mass distribution in the reduced space. The axes (D1 and D2) represent arbitrary dimensions from the dimensionality reduction process.}
        \label{fig:fig_bands_man}
\end{figure*}

\begin{figure*}
        \centering
        \includegraphics[width=0.9\linewidth]{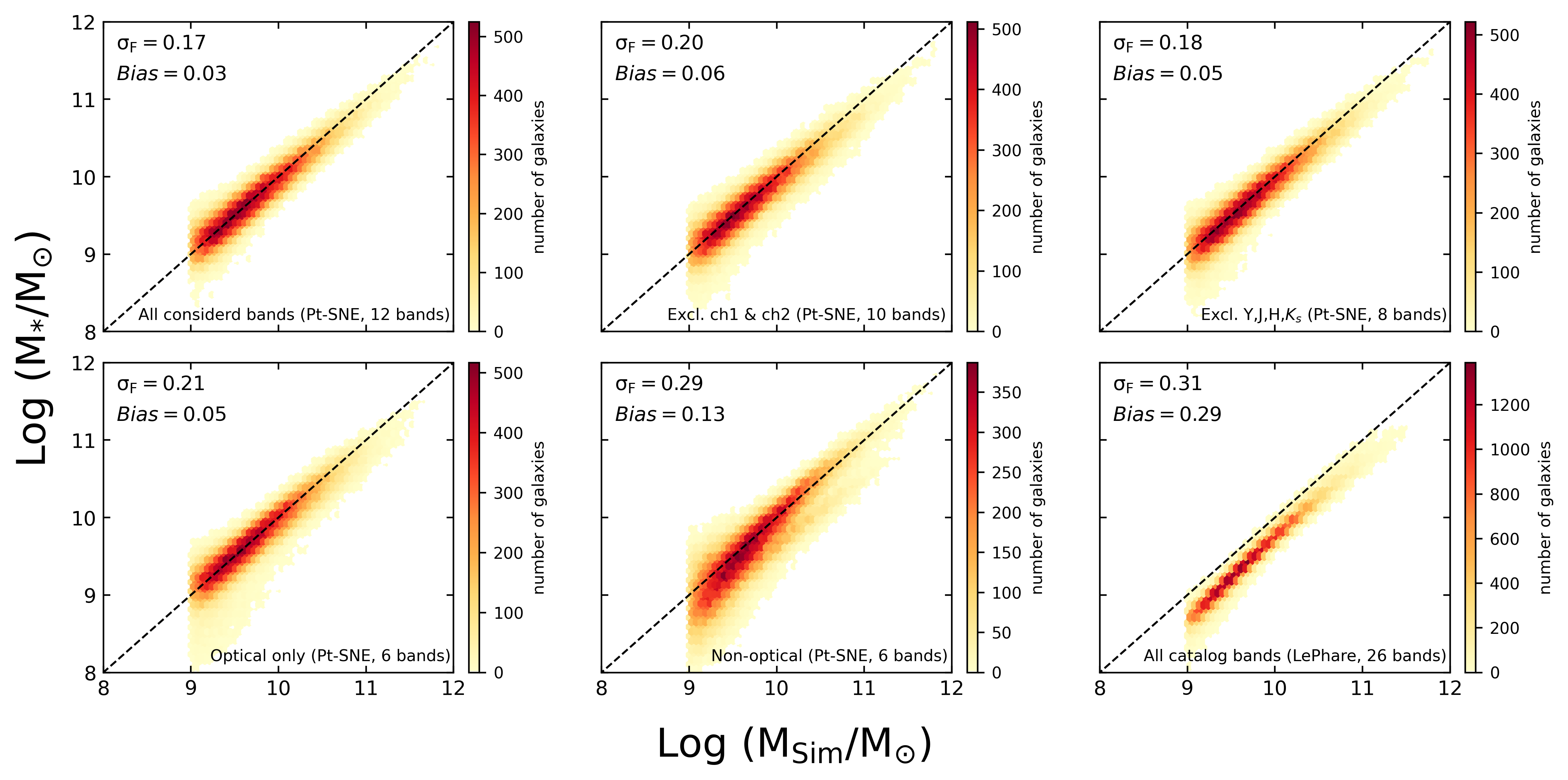}
        \caption{Performance comparison of stellar mass estimation between Pt-SNE (five band configurations listed in~\ref{tab:tab_acc_band}) and LePhare. Each panel shows the relationship between simulated masses ($M_{Sim}$) and estimated masses, with $\sigma_F$ (root-mean-square error) and bias quantifying the dispersion.}
        \label{fig:fig_acc_bands}
\end{figure*}

\subsection{Band-Dependence Performance}\label{sec:5.4}
We investigated how the number and type of photometric bands affect the performance of Pt-SNE, with results presented in Table~\ref{tab:tab_acc_band}. Pt-SNE was tested with five different band combinations, while LePhare used all 26 bands available in the catalog. Figure~\ref{fig:fig_bands_man} shows how different band configurations influence the manifold structure learned by Pt-SNE and the organization of mapped galaxies.

The analysis reveals that Pt-SNE achieves its best performance among the tested configurations (lowest errors across all metrics) when using all 12 available bands ($\sigma_F = 0.169$ dex, $\sigma_{NMAD} = 0.171$ dex). Performance degrades progressively as bands are removed, with the most significant impact occurring when using only non-optical bands (6 bands: $\sigma_F = 0.293$ dex), suggesting optical bands contain crucial information for mass estimation, likely due to their sensitivity to the Balmer and 4000 {\AA} breaks.

 Figure~\ref{fig:fig_acc_bands} provides a quantitative comparison of the estimation accuracy across different band configurations. Notably, excluding the ch1 and ch2 bands (10-band case) increases Pt-SNE's $\sigma_F$ by 16\% compared to the full 12-band case, while using only optical bands (6 bands) results in a 23\% increase. The outlier fraction shows particularly strong dependence, with OLF increasing 6.5-fold from the 12-band to optical-only case (0.004 to 0.026 dex).

Pt-SNE maintains superior performance over LePhare even in restricted band configurations. For instance, Pt-SNE's optical-only configuration ($\sigma_F = 0.208$ dex) outperforms LePhare's full 26-band results ($\sigma_F = 0.306$ dex). The normalized median absolute deviation shows similar trends, with Pt-SNE's worst band configuration (non-optical only: $\sigma_{NMAD} = 0.257$ dex) still substantially better than LePhare's result ($\sigma_{NMAD} = 0.415$ dex).

The results demonstrate that while Pt-SNE's accuracy is somewhat reduced with fewer bands, it maintains robust performance across all tested configurations, consistently outperforming LePhare's comprehensive 26-band analysis. This stark contrast underscores a major advantage of Pt-SNE: its remarkable efficiency in utilizing photometric information. It achieves superior stellar mass estimates with significantly fewer bands than LePhare.

\subsection{Computational Performance Comparison}\label{sec:5.5}
After evaluating the accuracy of Pt-SNE and LePhare, we compare their computational efficiency using a synthetic dataset of up to 10 million galaxies, generated by replicating our initial sample. All tests were performed on a machine equipped with an 11th Gen Intel\textregistered\ Core\texttrademark\ i7-1165G7 processor.

\begin{figure}
        \centering
        \includegraphics[width=0.95\linewidth]{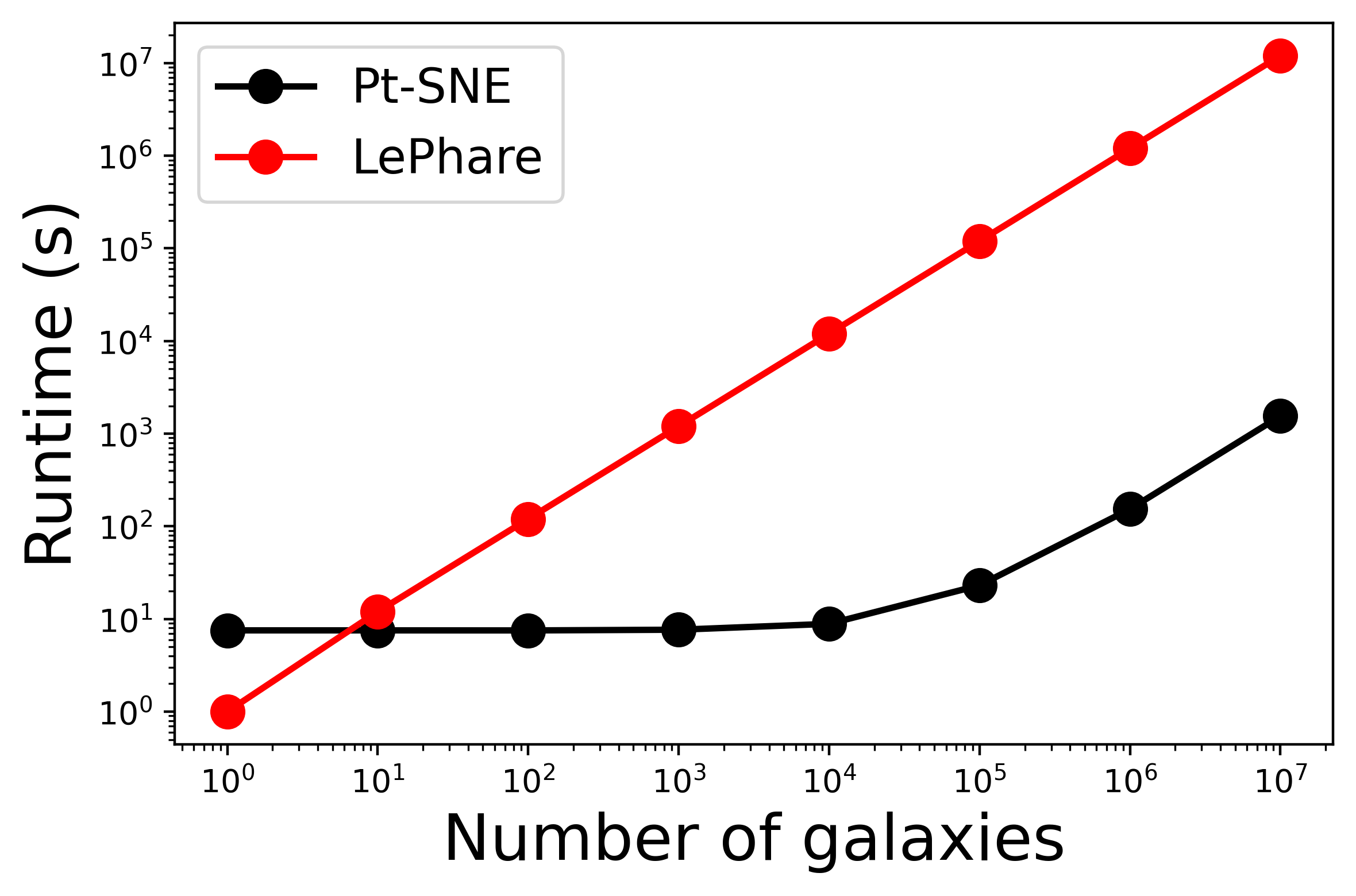}
        \caption{Runtime comparison between Pt-SNE and LePhare for galaxy sample sizes spanning $10^0$-$10^7$.}
        \label{fig:fig_ML_SED_com}
\end{figure}

Figure~\ref{fig:fig_ML_SED_com} illustrates the runtime scaling of both methods. For Pt-SNE, the reported runtime encompasses the full pipeline for processing new galaxy data, including: 1) training the manifold with BC03 synthetic galaxies, 2) mapping the mock galaxies onto the trained manifold, and 3) identifying the nearest neighbors in the manifold for subsequent stellar mass estimation (see Section~\ref{sec:3.4}). 

LePhare's runtime scales nearly linearly ($\mathcal{O}(n)$) with dataset size, as expected for template-fitting methods. In contrast, Pt-SNE exhibits a characteristic scaling curve: for a very small number of objects ($n \lesssim 10$), the runtime is dominated by the fixed cost of training, making it slower than LePhare for individual galaxies. However, for larger datasets, Pt-SNE exhibits sub-linear scaling ($\mathcal{O}(n \log n)$) and approaches linear complexity for larger samples ($\sim10^5$ to $10^7$ galaxies).

The speedup factor grows with sample size, reaching approximately $3.2 \times 10^{3}$ times faster than LePhare on average. This scaling relation is consistent with the computational performance of Pt-SNE reported in~\cite{asadi2025leveraging}. For the largest dataset (10$^7$ galaxies), Pt-SNE completes in $\sim$26 minutes while LePhare would require $\sim$139 days. This dramatic difference highlights Pt-SNE's suitability for large-scale analyses where traditional methods become computationally prohibitive.

\subsection{Training Set Dependence}\label{sec:5.6}
To quantitatively assess the sensitivity of the ML method to the representativeness of the training data, we conducted a controlled experiment evaluating how Pt-SNE performs when applied to galaxy populations outside its training domain. Since our method learns the mapping from colors to mass-to-light ratio ($M/L_{K_s}$), we partitioned the BC03 training set at the median $M/L_{K_s}$ value ($M/L_{K_s} = 0.13$) and trained separate models on the high and low $M/L_{K_s}$ subsets.

The results reveal a critical asymmetry in the model's extrapolation capability. When trained exclusively on high $M/L_{K_s}$ models (old stellar populations) and applied to our full simulation galaxy sample, Pt-SNE maintained its performance ($\sigma_{F} = 0.17$ dex, $\sigma_{STD} = 0.16$ dex, bias $= 0.06$ dex), demonstrating a robust ability to generalize to younger galaxy populations.

\begin{figure*}
        \centering
        \includegraphics[width=\linewidth]{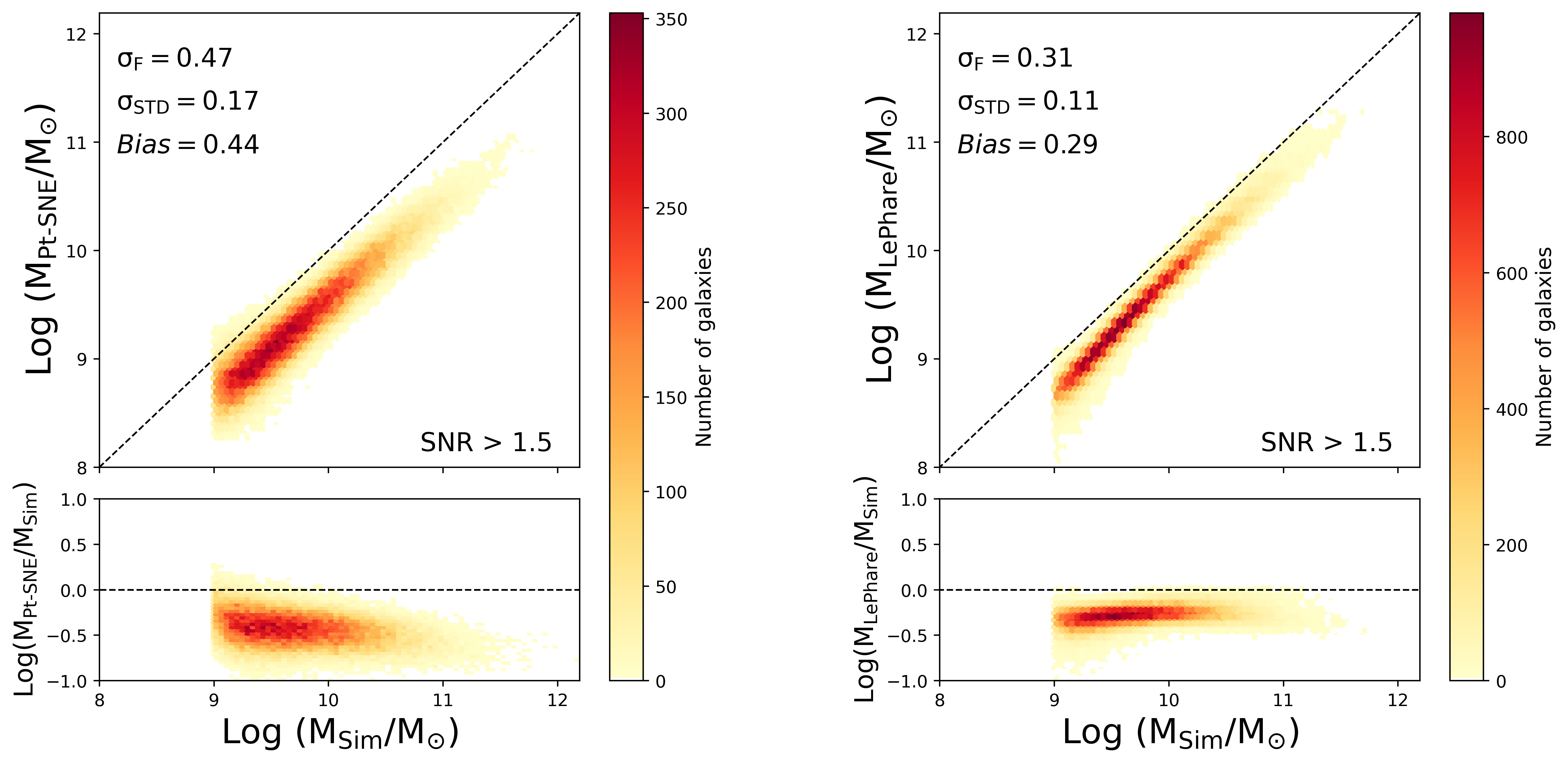}
        \caption{This figure compares stellar masses from a Pt-SNE model trained exclusively on low mass-to-light ratio populations ($M/L_{K_s} < 0.13$, the median value) and LePhare, with the true simulation values. $\sigma_F$, $\sigma_{STD}$ and Bias are root-mean-square, standard deviation and mean offset respectively.}
        \label{fig:fig_ML_lowvsLePh}
\end{figure*}

In contrast, the model trained only on low $M/L_{K_s}$ models (young stellar populations) exhibited a significant degradation when applied to the full simulation galaxy sample. While the random scatter remained low ($\sigma_{STD} = 0.17$ dex), a large systematic bias emerged (bias $= 0.44$ dex), leading to a high total error ($\sigma_{F} = 0.47$ dex). This pattern indicates that the model, when extrapolating beyond its training domain, can provide precise but systematically incorrect estimates (see Figure~\ref{fig:fig_ML_lowvsLePh}).

This experiment underscores a fundamental constraint of ML methods: their reliability is intrinsically bounded by the diversity of the training data. A model may fail catastrophically for galaxy populations not represented during training, producing inaccurate results. This highlights a disadvantage compared to SED-fitting method, which can in principle fit any galaxy within its template library's parameter space.

\section{Discussion}\label{sec:6}
This study presents a systematic comparison between machine learning (Pt-SNE) and traditional SED-fitting (LePhare) approaches for stellar mass estimation, using a COSMOS-like galaxy catalog derived from the Horizon-AGN hydrodynamical simulation as a benchmark with known true masses. By employing noise-injected BC03 synthetic models with 66 unique colors derived from 12 photometric bands for training Pt-SNE, we have established a controlled framework for evaluating the relative performance of these fundamentally different methodologies. Our study demonstrates that Pt-SNE significantly outperforms LePhare in the mass estimation of galaxies in both accuracy and execution time.

\subsection{Accuracy Comparison: The Bias Discrepancy}
The superiority of Pt-SNE in accuracy is primarily driven by its ability to avoid the substantial bias that plagues the SED-fitting method. The key finding of our controlled experiment—where both methods had access to the same underlying BC03 physics—is the large, mass-dependent bias in LePhare's estimates (0.286~dex) compared to the near-zero bias of Pt-SNE (0.029~dex).

While one would theoretically expect minimal bias when fitting BC03-generated galaxies with BC03 templates, we find that the observed bias can stem from fundamental methodological limitations rather than implementation errors. We identify three methodological limitations in LePhare with the potential to explain this systematic bias:

\begin{itemize}
    \item [--] The Horizon-AGN galaxies exhibit complex, merger-driven SFHs, while LePhare's template library is restricted to simple parametric forms (exponentially declining or delayed). This is analogous to reconstructing a complex symphony using only simple melodies—the best fit inevitably misses crucial nuances, leading to systematic error.

    \item [--] LePhare operates on a discrete, sparse grid of parameters (ages, metallicities, $\tau$). The true physical parameters of a simulated galaxy form a continuous distribution that will almost never coincide exactly with a grid point. During $\chi^2$ minimization, the algorithm is forced to select the nearest available template in this discrete space. This process introduces a systematic discretization error or quantization bias. Crucially, this error is not random; it depends on the location of the true parameters relative to the fixed grid. If the true parameter distribution is not uniformly centered within grid cells across the parameter space, the average of many such fits will exhibit a net bias. This is a known source of systematic error in grid-based Bayesian inference methods \citep[e.g., ][]{brammer2008eazy}. In principle, this grid-induced bias can be reduced by employing a significantly finer parameter grid. However, this mitigation strategy comes at a steep computational cost, as the number of templates—and thus the fitting time—scales with the product of the number of discrete values for each parameter. For large-scale surveys with millions of galaxies, this often makes an exhaustively fine grid computationally prohibitive, presenting a fundamental trade-off between accuracy and practical feasibility.
    
    \item [--] Even with identical underlying models (BC03), the notorious age-dust-metallicity degeneracy creates extended, near-flat valleys in the multi-dimensional $\chi^2$ surface. Within these valleys, many different parameter combinations yield similarly good fits to the photometry. In such a high-dimensional, degenerate likelihood landscape, the $\chi^2$ minimization (or Bayesian marginalization) does not converge to a unique solution centered on the truth. Instead, it can settle into any point within the degenerate valley. The final selected parameters are highly sensitive to the specific topography of the likelihood surface, noise realizations, and prior assumptions. This process systematically favors certain regions of parameter space (e.g., younger, dustier, and less massive solutions) over others, leading to a consistent bias in the ensemble of recovered masses \citep[e.g., ][]{conroy2009propagation,Pacifici2015}.
\end{itemize}

Further support for a methodological origin comes from the mass-dependence of the bias (Section~\ref{sec:5.3}). The bias is not a simple, constant offset; it is significantly worse at low masses. This trend is consistent with the known challenge of SED-fitting for low-mass galaxies, where the photometry is noisier and the age-dust-metallicity degeneracies are more pronounced. If the bias were solely due to a simple configuration error, such as a photometric zero-point offset, we would expect a more uniform bias across all stellar masses. The observed non-uniform trend suggests the bias is tied to the physical properties of the galaxies themselves and the limitations of the fitting process.

Most compellingly, our band-dependence analysis (Section~\ref{sec:5.4}) provides direct empirical proof of methodological inefficiency. Pt-SNE, utilizing only 6 optical bands, outperforms LePhare, which was provided all 26 available bands. If the bias were due to a simple configuration error, giving LePhare a 4-fold informational advantage should have mitigated the error. The fact that it did not indicates that LePhare's template-fitting approach is fundamentally less efficient at breaking degeneracies and extracting the stellar mass signal from the photometry compared to the machine learning model.

The significant difference in accuracy and robustness between the two methods directly highlights the fundamental methodological advantages of the Pt-SNE approach over traditional template fitting. Specifically, Pt-SNE avoids the limitations of LePhare due to several possible factors:

\begin{itemize}
    \item [--] ML algorithms excel at capturing complex, non-linear relationships between galaxy photometry and physical properties. Unlike SED-fitting, which is constrained to a specific family of physical models within its template library, ML methods can learn more flexible mappings from the data without being bound by pre-defined astrophysical parameterizations \citep[e.g.,][]{baron2019machine,kembhavi2022machine}.
    
    \item [--] ML methods can break degeneracies by learning in high-dimensional photometric space. Where SED-fitting might struggle to distinguish between, for example, young dusty galaxies and old metal-rich systems based on similar broad-band filters \citep[e.g.,][]{leja2017deriving,leja2019older}, an ML model can identify a unique solution by recognizing subtle, correlated patterns across the entire set of photometric bands simultaneously. This high-dimensional mapping allows it to separate galaxy types that appear degenerate in lower-dimensional projections.
    
    \item [--] ML approaches effectively use informative features, such as color indices, which are inherently differential. This approach provides a significant practical advantage: it makes the stellar mass estimate robust against systematic errors that affect all photometric bands similarly (e.g., global flux calibration issues or common-mode zero-point offsets). Unlike the $\chi^2$ minimization of template fitting, which is highly sensitive to absolute flux calibration (a sensitivity that can introduce systematic bias, as observed in LePhare), the ML models learn to prioritize the shape of the SED over the absolute flux scale, which contributes directly to its lower systematic bias.
\end{itemize}

The band-dependence analysis (Section~\ref{sec:5.4}) further highlights Pt-SNE's robustness: it achieves optimal performance with all 12 available bands yet still outperforms LePhare's 26-band results even when restricted to optical-only configurations (6 bands). This stark contrast underscores a major advantage of Pt-SNE: its remarkable efficiency in utilizing photometric information. It achieves superior stellar mass estimates with significantly fewer bands than a traditional SED-fitting method.

Beyond this general comparison, the mass-dependent performance of LePhare and Pt-SNE reveals specific limitations at both low and high stellar masses, resulting in increased errors compared to intermediate-mass galaxies—particularly for LePhare.

For low-mass galaxies, fainter intrinsic luminosities make observational noise and detection limits more significant factors in flux measurements. As demonstrated in Section~\ref{sec:5.3}, increasing the SNR by a factor of 100 reduces the errors for both methods in this regime. LePhare's template fitting approach is sensitive to these perturbations, possibly struggling with the degeneracy between dust attenuation and age. This degeneracy makes it difficult to distinguish between different physical scenarios that produce similar broadband filters, ultimately leading to larger uncertainties in stellar mass estimates. While Pt-SNE performs substantially better than LePhare, the fundamental challenges of limited photometric information and strong intrinsic parameter correlations at low SNR still contribute to slightly increased errors.

At high masses, increasing the SNR does not significantly improve either method's performance, indicating that observational noise is no longer the dominant source of uncertainty (see Figure~\ref{fig:fig_mass_bin}). LePhare's accuracy deteriorates markedly in this regime due to a fundamental mismatch between its simplified stellar population assumptions and the complex physics of massive galaxies in the HORIZON-AGN simulation. LePhare assumes exponentially declining SFHs, while the mock catalog contains massive galaxies with diverse SFHs shaped by mergers, AGN feedback, and rejuvenation events \citep{dubois2014dancing}. Additionally, the assumed restricted metallicity options in the mock catalog's existing LePhare configuration \citep{laigle2019horizon} fail to fully capture broader metallicity ranges characteristic of massive galaxies \citep{kaviraj2017horizon}. Subtle degeneracies among age, metallicity, and dust become more prominent in these evolved, older stellar populations, which LePhare's predefined templates struggle to disentangle, leading to less precise mass estimates. While Pt-SNE shows better performance, it is not immune to these limitations—having been trained on simple BC03 synthetic templates. This means Pt-SNE's current accuracy is also constrained by the realism and diversity of its own training data. Future work could explore the impact of training Pt-SNE on more physically diverse and complex SFHs and metallicity distributions.

\subsection{Computational Efficiency and Survey Scalability}
Regarding execution time, LePhare is significantly slower than Pt-SNE. This is because LePhare analyzes each galaxy in our mock sample independently, comparing its spectrum to a vast library of pre-computed model galaxies through an iterative fitting process. This individual analysis and extensive comparison make it computationally intensive. In contrast, Pt-SNE requires only an initial fast training phase on the BC03 synthetic templates; afterward, the entire mock sample can be rapidly mapped onto its manifold, enabling efficient stellar mass estimation. By avoiding individual galaxy analysis and iterative fitting, Pt-SNE offers a significantly faster approach, making it more suitable for large-scale applications where speed is crucial \citep[e.g.,][]{Hemmati2019,asadi2025leveraging}.

\subsection{Machine Learning Limitations abd Scope}
For clarity and reproducibility, this study focused exclusively on Pt-SNE as our ML benchmark, providing a controlled comparison against the established LePhare SED-fitting pipeline. While other ML approaches---including manifold learning algorithms (e.g., Self-Organizing Maps, UMAP~\citep{mcinnes2018umap}), regression methods (e.g., Random Forest, KNN~\citep{peterson2009k}), and clustering techniques (e.g., HDBSCAN~\citep{mcinnes2017hdbscan}, K-means)---could achieve comparable performance with similar data-driven advantages, our single-algorithm framework ensures unambiguous interpretation of results.

A notable limitation of ML approaches like Pt-SNE is their reduced interpretability compared to traditional SED-fitting. While SED-fitting explicitly links observed photometry to physical parameters (e.g., star formation history, metallicity, dust attenuation) through well-defined stellar population models, ML methods operate as ``black boxes", learning complex mappings without immediately transparent physical explanations for individual parameter derivations \citep[e.g.,][]{meher2021deep,sen2022astronomical}.

Another critical limitation is the dependence on the representativeness of the training data, as quantified in Section~\ref{sec:5.6}. While Pt-SNE demonstrates superior performance over LePhare in this work, its accuracy is ultimately constrained by the realism and diversity of the BC03 synthetic templates on which it was trained. The model's predictive power is therefore bounded by the physical parameter space covered by its training set.


\section{Conclusion}\label{sec:7}
This study compared the performance of machine learning (Pt-SNE) and traditional SED-fitting (LePhare) methods for estimating stellar masses of COSMOS-like galaxies derived from the Horizon-AGN simulation with known true masses. Pt-SNE was trained on BC03 synthetic models, while LePhare fits were derived from the same underlying physics, enabling a direct comparison of their methodological performance. The key findings are:

\begin{itemize}
    \item [--] Pt-SNE achieves superior accuracy, with a root-mean-square error ($\sigma_F$, Equation~\ref{eq:rms}) of 0.169 dex compared to LePhare's 0.306 dex. This improvement is primarily driven by Pt-SNE's significantly lower bias (0.029 dex versus LePhare's 0.286 dex, Equation~\ref{eq:bias}). This capacity for near-zero bias is a key advantage of the ML approach.
    
    \item [--] $\sigma_F$ and bias vary with stellar mass for both methods, but Pt-SNE exhibits much more stable performance and better accuracy across all mass ranges (from 9 to 12 in 0.5 dex bins). LePhare shows a U-shaped profile in both $\sigma_F$ and bias, with severe underestimation at both low- and high-mass extremes (0.33--0.36~dex bias) that diminishes at intermediate masses (0.24~dex). In contrast, Pt-SNE maintains consistently low bias ($<$0.10~dex) across most of the mass ranges, demonstrating its robustness.
    
    \item [--] Even with limited photometric data (6 optical bands), Pt-SNE achieves better accuracy than LePhare using all 26 available bands ($\sigma_F = 0.208$ vs. 0.306 dex). This superior performance stems directly from its much lower bias (0.053 dex vs. 0.286 dex), demonstrating that Pt-SNE's efficiency in extracting information translates into fundamentally more reliable mass estimates.
    
    \item [--] Pt-SNE shows remarkable computational efficiency, being $3.2 \times 10^{3}$ times faster than LePhare on average.
    
\end{itemize}

These results demonstrate that ML approaches can provide more accurate, robust, and computationally efficient stellar mass estimates than traditional SED-fitting. It is important to note that this study was conducted under idealized conditions where the stellar population models are known. When applied to real observational data, where the BC03 templates inevitably differ from actual galaxy spectra, both methods will be affected. However, the demonstrated methodological advantages of ML make it a highly promising tool for large galaxy surveys where both accuracy and processing speed are critical.

\section*{Acknowledgments}
We thank the anonymous referee for their constructive comments and suggestions that significantly improved the quality of this manuscript. We also extend our gratitude to Dr. Ghassem Gozaliasl (Department of Computer Science, Aalto University, PO Box 15400, Espoo 00 076, Finland) for his insightful discussions and valuable feedback during the preparation of this article.

\bibliography{ref}

\end{document}